\documentclass[11pt, a4paper, logo, copyright]{googlecloud}

\pdftrailerid{redacted}

\makeatletter
\renewcommand\bibentry[1]{\nocitep{#1}{\frenchspacing\@nameuse{BR@r@#1\@extra@b@citeb}}}
\makeatother

\usepackage{kantlipsum, lipsum}
\usepackage{dsfont}
\usepackage[authoryear, sort&compress, round]{natbib}

\usepackage[utf8]{inputenc} %
\usepackage[T1]{fontenc}    %
\usepackage{hyperref}       %
\hypersetup{
  colorlinks,
  citecolor=blue,
  linkcolor=red,
  urlcolor=blue}
\usepackage{booktabs}       %
\usepackage{nicefrac}       %
\usepackage{microtype}      %
\usepackage{wrapfig} %
\usepackage{soul} %
\usepackage{tikz, adjustbox}
\usepackage{arydshln}
\usepackage[capitalize,noabbrev]{cleveref}
\usepackage{fancyvrb}

\usepackage{inconsolata}
\usepackage[skins,breakable,most]{tcolorbox}

\usepackage{caption}
\usepackage{subcaption}
\usepackage{enumitem}
\usepackage{flushend}
\usepackage{balance}
\usepackage{lineno}
\usepackage{amsmath,amssymb,amsfonts}
\usepackage{algorithm}
\usepackage{algpseudocode}
\usepackage{graphicx}
\usepackage{textcomp}
\usepackage{xcolor}
\usepackage{xspace}
\usepackage{colortbl}
\usepackage{amsthm}
\usepackage{url}
\usepackage{float}
\usepackage{multirow}
\usepackage{multicol}
\usepackage{color}
\usepackage{bm}
\usepackage{bbm}
\usepackage{lmodern}
\usepackage{threeparttable}
\usepackage{wrapfig}

\usepackage{pifont}

\usepackage{array}
\newcolumntype{L}[1]{>{\raggedright\let\newline\\\arraybackslash\hspace{0pt}}m{#1}}
\newcolumntype{C}[1]{>{\centering\let\newline  \\\arraybackslash\hspace{0pt}}m{#1}}
\newcolumntype{R}[1]{>{\raggedleft\let\newline \\\arraybackslash\hspace{0pt}}m{#1}}

\definecolor{PromptBackground}{HTML}{F8F8F8}
\definecolor{PromptFrame}{HTML}{D0D0D0}
\definecolor{PromptTitle}{HTML}{2A628F}
\definecolor{PromptSection}{HTML}{2A628F}
\definecolor{PromptVariable}{HTML}{C93C00}
\definecolor{PromptComment}{HTML}{BEBEBE}
\definecolor{CodeBackground}{HTML}{FDFDFD}

\newcommand{\var}[1]{\textcolor{PromptSection}{\texttt{#1}}}
\newcommand{\inputvar}[1]{\textcolor{PromptVariable}{\texttt{#1}}}

\lstdefinestyle{mypython}{
    language=Python,
    backgroundcolor=\color{CodeBackground},
    basicstyle=\ttfamily\small,
    keywordstyle=\color{Blue}\bfseries,
    stringstyle=\color{Green},
    commentstyle=\color{Gray}\itshape,
    identifierstyle=\color{Black},
    numbers=none,
    frame=tb, 
    framerule=1pt,
    rulecolor=\color{CodeBackground},
    framesep=5pt,
    breaklines=true,
    keepspaces=true,
    showstringspaces=false,
    literate= 
      {_}{\_}{1}
      {\{}{{\{}}{1}
      {\}}{{\}}}{1}
      {"}{"} {1}
      {'}{'} {1}
}

\newcommand{\ours}{\textsc{VeriGuard}\xspace}
\newcommand{\quotes}[1]{``#1''}

\definecolor{lightorange}{RGB}{245, 237, 211}
\definecolor{clovergreen}{RGB}{32,115,55}

\newtcbox{\hlprimarytab}{on line, rounded corners, box align=base, colback=c3!10,colframe=white,size=fbox,arc=3pt, before upper=\strut, top=-2pt, bottom=-4pt, left=-2pt, right=-2pt, boxrule=0pt}
\newtcbox{\hlsecondarytab}{on line, box align=base, colback=blue!10,colframe=white,size=fbox,arc=3pt, before upper=\strut, top=-2pt, bottom=-4pt, left=-2pt, right=-2pt, boxrule=0pt}
\newtcbox{\hlcasetab}{on line, box align=base, colback=c5!10,colframe=white,size=fbox,arc=3pt, before upper=\strut, top=-2pt, bottom=-4pt, left=-2pt, right=-2pt, boxrule=0pt}

\definecolor{c1}{cmyk}{0,0.6175,0.8848,0.1490} 
\definecolor{c2}{cmyk}{0.1127,0.6690,0,0.4431} 
\definecolor{c3}{cmyk}{0.3081,0,0.7209,0.3255} 
\definecolor{c4}{cmyk}{0.6765,0.2017,0,0.0667} 
\definecolor{c5}{cmyk}{0,0.8765,0.7099,0.3647}

\usepackage[utf8]{inputenc} 
\usepackage[T1]{fontenc}    
\usepackage{amsfonts}       
\usepackage{comment} 
\usepackage{mdframed}
\usepackage{fontawesome}
\usepackage{csquotes}
\usepackage{makecell}
\definecolor{beigecolor}{RGB}{253, 244, 204} 
\definecolor{greencolor}{RGB}{228, 242, 217} 
\definecolor{bluecolor}{RGB}{66, 133, 244} 
\definecolor{orgcolor}{RGB}{255, 140, 15} 

\definecolor{redcolor}{RGB}{234, 67, 53} 

\definecolor{ggreen}{RGB}{52, 168, 83}

\definecolor{gyellow}{RGB}{251, 188, 5}

\definecolor{lightorange}{RGB}{245, 237, 211} 
\definecolor{bluebar}{RGB}{138,159,201}
\definecolor{pinkbar}{RGB}{232,180,189}

\usepackage{mathtools}

\lstdefinestyle{mystyle}{
    backgroundcolor=\color{backcolour},   
    commentstyle=\color{codegreen},
    keywordstyle=\color{magenta},
    numberstyle=\tiny\color{codegray},
    stringstyle=\color{codepurple},
    basicstyle=\ttfamily\scriptsize,
    breakatwhitespace=false,         
    breaklines=true,                 
    captionpos=b,                    
    keepspaces=true,                 
    numbers=left,                    
    numbersep=5pt,                  
    showspaces=false,                
    showstringspaces=false,
    showtabs=false,                  
    tabsize=2,
    frame=none,
    aboveskip=1pt,
    belowskip=1pt,
}
\lstdefinestyle{plainins}{
    backgroundcolor=\color{white},   
    commentstyle=\color{codegreen},
    keywordstyle=\color{magenta},
    numberstyle=\tiny\color{codegray},
    stringstyle=\color{codepurple},
    basicstyle=\ttfamily\scriptsize,
    breakatwhitespace=false,         
    breaklines=true,                 
    captionpos=b,                    
    keepspaces=true,                 
    numbers=none,                    
    numbersep=5pt,                  
    showspaces=false,                
    showstringspaces=false,
    showtabs=false,                  
    tabsize=2,
    aboveskip=0pt,
    belowskip=0pt,
    frame=single
}
\lstdefinestyle{plainexam}{
    backgroundcolor=\color[HTML]{FFFCF3},   
    commentstyle=\color{codegreen},
    keywordstyle=\color{magenta},
    numberstyle=\tiny\color{codegray},
    stringstyle=\color{codepurple},
    basicstyle=\ttfamily\scriptsize,
    breakatwhitespace=false,         
    breaklines=true,                 
    captionpos=b,                    
    keepspaces=true,                 
    numbers=none,                    
    numbersep=5pt,                  
    showspaces=false,                
    showstringspaces=false,
    showtabs=false,                  
    tabsize=2,
    aboveskip=0pt,
    belowskip=0pt
}

\title{VeriGuard: Enhancing LLM Agent Safety via Verified Code Generation}

\correspondingauthor{Corresponding authors:  Lesly Miculicich lmiculicich@google.com and Long T. Le: longtle@google.com}

\renewcommand{\today}{}

\author[1]{Lesly Miculicich}
\author[1]{Mihir Parmar}
\author[1]{Hamid Palangi}
\author[2]{Krishnamurthy Dj Dvijotham}
\author[3]{Mirko Montanari}
\author[1*]{Tomas Pfister}
\author[1*]{Long T. Le}

\affil[1]{Google Cloud AI Research}
\affil[2]{Google DeepMind}
\affil[3]{Google Cloud AI}

\begin{abstract}

The deployment of autonomous AI agents in sensitive domains, such as healthcare, introduces critical risks to safety, security, and privacy. These agents may deviate from user objectives, violate data handling policies, or be compromised by adversarial attacks. Mitigating these dangers necessitates a mechanism to formally guarantee that an agent's actions adhere to predefined safety constraints, a challenge that existing systems do not fully address.
We introduce \ours{}, a novel framework that provides formal safety guarantees for LLM-based agents through a dual-stage architecture designed for robust and verifiable correctness. The initial offline stage involves a comprehensive validation process. 
It begins by clarifying user intent to establish precise safety specifications. \ours{} then synthesizes a behavioral policy and subjects it to both testing and formal verification to prove its compliance with these specifications. 
This iterative process refines the policy until it is deemed correct. Subsequently, the second stage provides online action monitoring, where \ours{} operates as a runtime monitor to validate each proposed agent action against the pre-verified policy before execution. This separation of the exhaustive offline validation from the lightweight online monitoring allows formal guarantees to be practically applied, providing a robust safeguard that substantially improves the trustworthiness of LLM agents.
\end{abstract}

\begin{document}

\maketitle

\section{Introduction}

The proliferation of Large Language Model (LLM) agents marks a significant leap towards autonomous AI systems capable of executing complex, multi-step tasks \citep{xi2023rise, yao2022react}. These agents, often empowered to interact with external tools, APIs, and file systems \citep{schick2023toolformer, patil2023gorilla}, hold immense promise for automating digital workflows and solving real-world problems. However, this power introduces substantial and often unpredictable safety and security vulnerabilities. A critical reliability gap has emerged: while LLM agents can generate solutions with unprecedented flexibility, the solution they produce often lacks assurances, making it susceptible to subtle errors, security flaws, and emergent behaviors that can lead to catastrophic failures. An agent tasked with data analysis could inadvertently exfiltrate sensitive information; one managing cloud infrastructure could execute destructive commands; another interacting with financial APIs could trigger erroneous, irreversible transactions. This problem is even more serious when there is adversary attack on the system, as shown in \cite{zhangASB}.

Existing safety mechanisms—such as sandboxing, input/output filtering, and static rule-based guardrails \citep{meta2024llama_guard, rebedea-etal-2023-nemo} —provide a necessary but insufficient first line of defense. These approaches are fundamentally reactive or based on pattern matching; they struggle to cover the vast and dynamic state space of agent actions and can be bypassed by novel adversarial inputs or unforeseen edge cases \citep{wei2023jailbroken, xu2023llmfoolitselfpromptbased}. They lack a deep, semantic understanding of the code's intent and consequences, treating the agent's output as a black box to be constrained. This leaves systems vulnerable to sophisticated exploits that a static rule set cannot anticipate \citep{Schulhoff2022ignore}. For LLM agents to be trusted in high-stakes, mission-critical environments, a more rigorous, provable approach to safety is required.

In this work, we propose a novel method to address this reliability gap, centered on the \ours{} framework. \ours represents a paradigm shift from reactive filtering to proactive, provable safety by deeply integrating policy specification generation and automated verification into the agent's action-generation pipeline. VeriGuard fundamentally reshapes the code generation process to be \quotes{correct-by-construction}. This is achieved by prompting the LLM agent to generate not only the functional code for an action but also its corresponding verification that precisely define the code's expected behavior and safety properties. These paired artifacts are then immediately subjected to an automated verification engine. An iterative refinement loop forms the core of our framework: if verification fails, the verifier provides a specific counterexample or logical inconsistency, which is fed back to the agent as a concrete, actionable critique to guide the generation of a corrected and verifiably safe version of the code \citep{pan-etal-2024-automatically, zhao2025recodeimprovingllmbasedcode}. More details are in \textsection \ref{sec: method}.

The primary contribution of this paper is the VeriGuard framework itself, which includes novel methodologies for the LLM-driven generation and refinement of verifiable code tailored to agent security and safety contexts. We further contribute a comprehensive empirical validation of the framework's effectiveness in preventing unsafe actions across a variety of challenging domains. Finally, we present a detailed analysis of the performance trade-offs inherent in this approach. 

\section{Related Work}
\label{sec: related_work}

\subsection{LLM Agents and the Emergence of Autonomous Systems}

The development of Large Language Models (LLMs) has catalyzed the emergence of a new class of autonomous systems known as LLM agents. LLM agents are designed to be proactive, goal-oriented entities capable of planning, reasoning, and interacting with their environment through the use of tools \citep{schick2024toolformer}. Early frameworks like ReAct demonstrated how to synergize reasoning and acting within LLMs, enabling them to solve complex tasks by generating both textual reasoning traces and executable actions \citep{yao2022react}. The agent can also execute more complex tasks like web browsing. This capability, however, is merely the entry point into a broader spectrum of autonomous actions. Advanced agents are not just navigating websites but are becoming generalist problem-solvers on the web and beyond. This evolution is detailed in research and demonstrated in benchmarks like WebArena \citep{zhou2023webarena} and Mind2Web \citep{gou2025mind2web2}, which test agents on their ability to perform multi-step, realistic tasks on live websites.

This paradigm quickly evolved into more sophisticated agent architectures. Systems like AutoGPT \citep{AutoGPT} and BabyAGI showcased the potential for fully autonomous task completion, where agents could decompose high-level goals into smaller, executable steps, manage memory, and self-direct their workflow. Further research has explored enhancing agent capabilities through mechanisms like self-reflection and verbal reinforcement learning, allowing them to learn from past mistakes and improve their performance over time \citep{shinn2023reflexion}. The concept of "Generative Agents" pushed the boundaries even further by creating interactive simulacra of human behavior within a sandbox environment, highlighting the potential for complex social and emergent behaviors \citep{park2023generative}. A comprehensive survey by \citep{wang2023survey} details the rapid advancements and architectural patterns in this burgeoning field.

\subsection{LLM Safety, Alignment, and Guardrails}

A significant body of research has focused on ensuring the safety and alignment of LLMs. A primary line of defense involves creating guardrails to constrain agent behavior. These can range from simple input/output filtering and prompt-based restrictions to more sophisticated techniques \citep{bai2022constitutional}. Another critical area is the proactive discovery of vulnerabilities through ``red teaming'', where humans or other AIs craft adversarial prompts to elicit unsafe or undesirable behaviors from the model \citep{perez2022red}. The insights from these attacks are then used to fine-tune the model for greater robustness. Despite these efforts, LLMs remain susceptible to a wide array of "jailbreaking" techniques that can bypass safety filters \citep{wei2023jailbroken}. More recent work has focused on creating safety-tuned LLMs specifically for tool use, aiming to prevent harmful API calls or command executions \citep{jin2024llm}.

There are some previous work in Agent safeguard. GuardAgent, a framework that uses an LLM-based ``guard agent'' to safeguard other LLM agents. GuardAgent operates as a protective layer, using reasoning to detect and prevent unsafe behaviors \citep{xiang2024guardagent}. Another work is ShieldAgent, a guardrail agent designed to ensure that autonomous agents powered by large language models (LLMs) adhere to safety policies \citep{chen2025shieldagent}.

However, these existing approaches are largely empirical and reactive. They rely on identifying and patching vulnerabilities as they are discovered, but they do not provide formal, provable guarantees of safety. A clever adversary can often devise a novel attack that circumvents existing guardrails. This highlights a fundamental limitation: without a formal specification of what constitutes ``safe'' behavior and a method to verify compliance, safety remains an ongoing. VeriGuard distinguishes itself from this body of work by moving from an empirical to a formal verification paradigm, aiming to prove the correctness of an agent's actions before they are ever executed.

\subsection{Formal Methods and Verifiable Code Generation}

Formal methods provide a mathematically rigorous set of techniques for the specification, development, and verification of software and hardware systems. The advent of powerful LLMs has opened a new frontier for bridging the gap between natural language specifications and formal, machine-checkable code. Recent research has begun to explore the potential for LLMs to automate or assist in the generation of not just code, but also its formal specification and verification artifacts. For example, \citep{li2024llmbased} demonstrate a system where LLMs are used to generate verifiable computation, producing code along with the necessary components for a verifier to check its correctness. Further studies have investigated the self-verification capabilities of LLMs \citep{ghaffarian2024can}. This line of work shows the promise of integrating LLMs into high-assurance software development pipelines.

\section{Methodology}\label{sec: method}

\begin{figure}[htbp]
    \centering 
    \vspace{-3mm}
    \includegraphics[width=1\textwidth]  
    {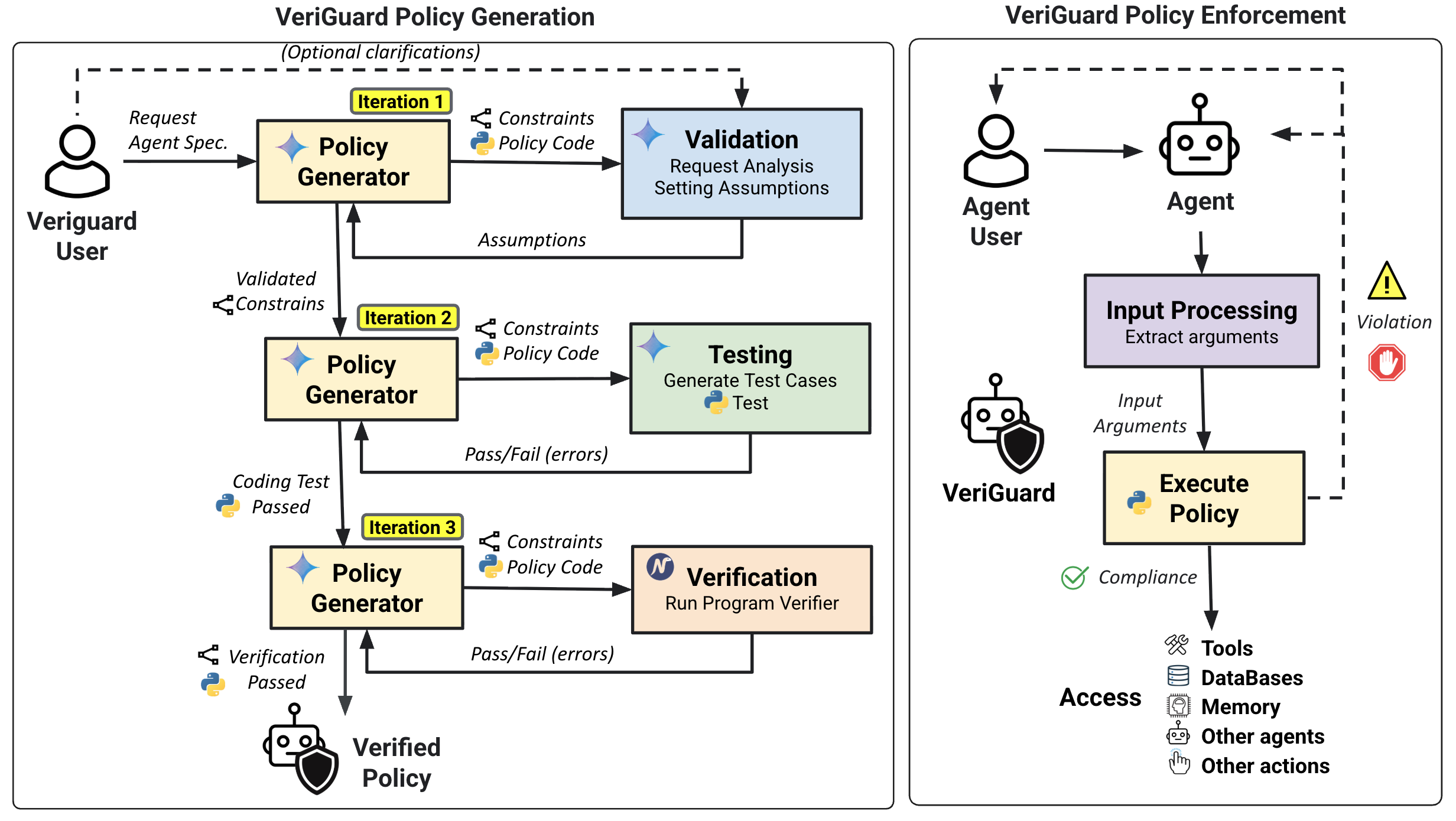} 
    \vspace{-5mm}
    \caption{\ours{} overview which includes Policy generation and Policy enforcement. The verified policy is integrated into the agent as a runtime safeguard, intercepting and preventing harmful actions. } 
    \label{fig:overview} 
\vspace{-3mm}
\end{figure}

Figure \ref{fig:overview} describes the high-level ideas of VeriGuard, which operates in two main stages: \textbf{(i) Policy Generation}: VeriGuard takes inputs including the agent's specification and a high-level security request in natural language to synthesize an initial policy function and its corresponding formal constraints. To ensure the correctness and alignment of this policy, we employ a rigorous, multi-step refinement feedback loop. This loop begins with a validation phase to resolve any ambiguities in the user's request, followed by an automated code testing phase that generates unit tests to verify functional correctness. The most critical phase uses formal verification to prove that the policy code adheres to its specified conditions, ensuring a provably-sound safety contract. \textbf{(ii) Policy Enforcement}: The verified policy is integrated into the agentic system at key enforcement points, where it intercepts and evaluates agent-initiated actions before execution. When a potential violation is detected, VeriGuard can employ one of several enforcement strategies, ranging from immediately terminating the agent's task to blocking the specific unsafe action or engaging in a collaborative re-planning dialogue with the agent. 

\subsection{Task Definition}

In this section, we formalize the process of generating agent policies from high-level, natural language specifications.

\paragraph{Policy Generation} 

Given a safety and security request in natural language, denoted as $r$, and a  agent specification, $\mathcal{S}$, the primary objective is to synthesize a policy function, $p$, written in a structured programming language. Concurrently, a set of verifiable constraints (i.e. pre and post-conditions), $C = \{c_1, c_2, \dots, c_n\}$, is derived. The system must guarantee that the generated policy $p$ complies with all constraints in $C$. This relationship is formally denoted as $p \models C$, signifying that $\forall c \in C$, the policy $p$ satisfies $c$. The user request $r$ typically defines a security or operational protocol in text format, while the agent specification $\mathcal{S}$ provides a schematic of the agent: task description, input/output (I/O) data structures, available context, environmental information, and any other available data.

\paragraph{Policy Enforcement} Given an agentic system and a set of verified policies, the second objective is to integrate these policies as enforcement mechanisms. The goal is to optimize the system's performance by minimizing policy violations (i.e., reducing the attack surface) while maximizing the agent's task-completion utility.

\subsection{Framework}
To address the defined tasks, we propose a framework, \textbf{\ours{}}, which consists of an initial policy generator followed by an iterative refinement loop. This loop validates, tests, and formally verifies the policy code to ensure it accurately reflects the agent requirements and specifications. For the policy enforcement task, experiment with multiple integration strategies for deploying \ours{} within an agentic system.

\subsubsection{Policy Generator}
The Policy Generator is the core component responsible for translating the agent's specification and user's intent into executable code and formal specifications. It has two sub-components: (1) policy code generation, and (2) constrains generation, both LLM-based. At the first pass, the policy generator takes the user request $r$, an agent specification $\mathcal{S}$, to produce a preliminary policy function $p_0$, together with a list of arguments the policy function requires $\mathcal{P}_{0}$. Similarly, the constrains generator take same inputs to produce a set of constraints $C_0$. This initial generation functions, $G_0$ and $H_0$, can be represented as:
$$G_0(r, \mathcal{S}) \rightarrow (p_0, \mathcal{P}_{0}) \qquad H_0(r,\mathcal{S}) \rightarrow (C_0)$$

The arguments schema $\mathcal{P}_{0}$ contains the name, description and type of each required input argument of the policy function. If the request entails multiple interdependent rules, the generator produces a single, cohesive codebase that encapsulates all logic. The prompts for the initial generations are detailed in \ref{ap:prompt_generation}.

The Policy Generator operates within an iterative refinement loop where  policy and constraints are gradually improved from the previous step ($p_{t-1}, C_{t-1}$):
$$G_t(r, \mathcal{S}, R, A, e, p_{t-1}) \rightarrow (p_t, \mathcal{P}_{t}) \qquad H_t(r, \mathcal{S}, R, A, p_{t-1}) \rightarrow (C_t)$$
$R$, $A$, and $e$ are the set of requirements, assumptions and coding error messages. 

\subsubsection{Refinement Process}
We employ a three-stage refinement process: validation, testing, and formal verification. 

\paragraph{Validation}
The Validator's primary function is to resolve ambiguities and ensure the semantic alignment between the user's natural language request and its formal representation ($p_0, C_0$). This process is bifurcated into an analysis phase and a disambiguation phase.

In the analysis phase, a function $V_a$ scrutinizes the initial artifacts to identify semantic ambiguities, logical inconsistencies, and implicit presuppositions. The output is a set of queries, $Q$, that encapsulate these issues: $V_a(p_0, C_0) \rightarrow Q$

In the disambiguation phase, a function $V_d$ processes the user's feedback, $U_{\text{feedback}}$, to resolve the queries in $Q$. This interactive process yields a definitive set of explicit assumptions, $A$, and a refined, unambiguous set of requirements $R$ as: $V_d(Q, U_{\text{feedback}}) \rightarrow (A, R)$

In an autonomous operational mode where user feedback is unavailable, an internal module, $\Omega$, is invoked to resolve the queries by selecting the most contextually plausible interpretations. This generates a set of default assumptions, $A_{\text{default}}$, which are then used to produce the final requirements $R$. This autonomous path is modeled as: $V_d(Q, \Omega(Q)) \rightarrow (A_{\text{default}}, R)$. \ref{ap:prompt_validation} shows the implementations detail of this component.

\paragraph{Code Testing}
This module automatically generates a suite of test cases to perform empirical validation of the policy function. It takes the policy code $p$, the user request $r$, and the agent specification $\mathcal{S}$ as input. The objective is to ensure that the policy meets a baseline of functional requirements and correctly handles typical and edge-case scenarios before proceeding to the more computationally expensive formal verification stage. The output is a set of test cases formatted for the \textit{PyTest} framework. The policy code is refined iteratively until all generated tests pass, with failure reports and error messages $e$ serving as feedback for the refinement loop. The iteration stops when not more errors are found or at a maximum $N$ number. Details in \ref{ap:prompt_testing}.

\paragraph{Verification}
The final stage of refinement involves formal verification using a program verifier. This component takes the logical constraints $C$ and the policy code $p$ as input. The constraints in $C$ define a formal contract, specifying the pre-conditions ($C_{\text{pre}} \subseteq C$) that must hold before the policy's execution and the post-conditions ($C_{\text{post}} \subseteq C$) that must be guaranteed upon its completion.

The verifier's task is to mathematically prove that the policy code $p$ adheres to this contract. This relationship is formally expressed using a Hoare triple: $\{C_{\text{pre}}\} \ p \ \{C_{\text{post}}\}$.
If program $p$ starts in a state where pre-condition $C_{\text{pre}}$ is true, its execution is guaranteed to terminate in a state where post-condition $C_{\text{post}}$ is true. If the code violates the contract, the verifier provides a counterexample or error trace $e$, which is used as feedback to refine the policy or constraints. The refinement cycle continues until formal verification succeeds or at a maximum $N$ number. For this implementation, we utilize the Nagini verifier \citep{eilers2018nagini} as a black box. As a static verifier built on the Viper \citep{EilersSchwerhoffSummersMueller25} infrastructure, Nagini can handle more complex properties than other available Python verifiers. Pre-processing for Nagini is detailed in \ref{ap:prompt_verification}. 

\subsection{Policy Enforcement Strategies}
Once a policy is generated and verified, it is integrated into the agentic system at specific enforcement points that intercept agent-initiated actions (e.g., tool executions, database access, environmental interactions). Each agent can be governed by one or more policy functions. 

\subsubsection{Policy Function Arguments}
At runtime, the arguments for the policy function defined, in $\mathcal{P}$, must be populated from the agentic system data defined in $S$. We do not assume $S$ is a direct input to the policy, as this data could be unstructured, and require preprocessing or extraction. Moreover, implementing preprocessing step strictly via code can limit the system's flexibility.
Thus, a function $f: S \to P$ is required to map the agent data to the policy arguments. For our experiments, we implement $f$ as a flexible LLM-based component (\ref{ap:prompt_enforce}). The input of $f$ is the agent data in the format specified in $S$ and the output is a populated dictionary of arguments specified in $P$.

\subsubsection{Policy Function Integration}\label{subsec:integration} 

We experimented with four distinct enforcement strategies upon detecting a policy violation: \textbf{(i) Task Termination:} the most restrictive approach, which halts the agent's entire high-level task and issues a notification explaining the violation; \textbf{(ii) Action Blocking:} a more targeted approach, where the specific action that violates the policy is blocked, but the agent is permitted to continue executing subsequent actions in its plan that do not violate policy; \textbf{(iii) Tool Execution Halt:} which stops the specific execution that caused the violation and returns no observation to the agent, forcing the agent's reasoning process to halt and decide on a new course of action based on the lack of feedback; and \textbf{(iv) Collaborative Re-planning:} the most interactive strategy, which informs the agent of the specific policy it has violated and requests that it generate a new action plan that complies with the stated constraints.

\section{Experiments}
\label{sec: exp}

\subsection{Datasets} \label{sec: exp-data}
\paragraph{Agent Security Bench (ASB)} \citep{zhangASB} 
This dataset focuses on four types of attacks against agentic systems: direct and indirect prompt injection, memory poisoning, and plan-of-thought backdoors. The framework utilizes 10 ReAct agents across 10 different scenarios (e.g., finance, commerce, autonomous driving), incorporating 10 different tasks, 13 attack variations, and 400 attack tools. 

\paragraph{EICU-AC} \citep{xiang2024guardagent} 
This benchmark evaluates an ICU question/answer agent's access control over 10 patient databases. It defines specific user roles, each restricted to designated subsets of the databases and their internal information categories. The agent must deny any user question if its answer requires accessing data that is forbidden for the user's given role.

\paragraph{Mind2Web-SC} \citep{xiang2024guardagent} 
Mind2Web-SC adapts the Mind2Web benchmark, which tasks web agents with solving over 2,000 complex online tasks using sequential actions. The \quotes{SC} (Safety Control) variant augments this framework by introducing safety rules that prohibit specific users from performing certain web activities, thereby testing the agent's ability to enforce these policies.

\subsection{Evaluation Metrics}

\paragraph{ASB}: we simultaneously evaluate defense efficacy and operational interference using two primary metrics: the \textbf{(i) Attack Success Rate (ASR$\downarrow$)}: quantifies the quality of the defense. It is defined as the ratio of successful attacks to the total number of attack attempts. 
\textbf{(ii) Task Success Rate (TSR$\uparrow$)}: measures the operational interference or overhead imposed by the defense mechanism. It is calculated as the ratio of successfully completed original tasks to the total number of tasks attempted.

\paragraph{EICU-AC} and \textbf{Mind2Web-SC}: the task is defined exclusively as access control, where correctly stopping a forbidden action is the desired outcome. This evaluation is treated as a binary classification problem (Allow vs. Deny), so we use standard classification metrics: \textbf{Accuracy}, \textbf{Precision}, and \textbf{Recall}.

\subsection{Baselines}
\paragraph{ASB}: we experiment with four backbone LLMs: Gemini 2.5 Flash and Pro, GPT-4.1, and Claude-Sonnet-4. We establish two base boundaries: the upper-bound \emph{No Attack}, derived from unattacked model performance, and \emph{No Defence}, derived from undefended model performance.
We compare against several baselines: \textbf{Paraphrasing} \citep{jain2023baselinedefensesadversarialattacks}, which rewords the query to disrupt malicious special-character sequences and triggers (effective for DPI and PoT Backdoor attacks); \textbf{Dynamic Prompt Rewriting} \citep{zhangASB}, which transforms the input to align with security objectives (proposed for DPI); and \textbf{Delimiter} \citep{mattern-etal-2023-membership}, which encapsulates the user query to ensure bounded execution (effective for IPI). We also implemented a stronger \textbf{Guardrail} baseline that receives the same input as \ours{} but, instead of generating a code function, directly asks an LLM to evaluate policy compliance.

\paragraph{EICU-AC} and \textbf{Mind2Web-SC}: we report the results of several state-of-the-art (SOTA) approaches. These include \textbf{GuardAgent} \citep{xiang2024guardagent}, which translates natural language safety rules into executable code via manually defined functions; \textbf{AGrail} \citep{luo-etal-2025-agrail}, which implements a mechanism to continually learn and adapt policies (as security checks) and uses an LLM for verification; \textbf{LLaMA-Guard 3} \cite{dubey2024llama3herdmodels}, a model trained to detect security issues; and \textbf{AgentMonitor} \citep{chan2024agentmonitorplugandplayframeworkpredictive}, a guardrail method for multi-agent systems. We also include the \textbf{Hard-coded Rules} baseline in  \citep{xiang2024guardagent}.

\begin{table}[!t]
\centering\setlength{\tabcolsep}{5.3pt}
\small
\caption{Experiment results of \ours{} on ASB benchmark. Attack Success Rate (ASR $\downarrow$) Task Success Rate (TSR  $\uparrow$).}
  \label{table: asb}
  \resizebox{\textwidth}{!}{%
  \begin{tabular}{lcccccccccc}
    \toprule
    \multirow{2}{*}{\textbf{Defense}} &\multicolumn{2}{c}{\textbf{DPI}}&\multicolumn{2}{c}{\textbf{IPI}}&\multicolumn{2}{c}{\textbf{MP}}&\multicolumn{2}{c}{\textbf{PoT}}&\multicolumn{2}{c}{\textbf{AVG}}\\
\cmidrule(lr){2-3}\cmidrule(lr){4-5}\cmidrule(lr){6-7}\cmidrule(lr){8-9}\cmidrule(lr){10-11}
&\textit{ASR$\downarrow$}&\textit{TSR$\uparrow$}&\textit{ASR$\downarrow$}&\textit{TSR$\uparrow$}&\textit{ASR$\downarrow$}&\textit{TSR$\uparrow$}&\textit{ASR$\downarrow$}&\textit{TSR$\uparrow$}&\textit{ASR$\downarrow$}&\textit{TSR$\uparrow$} \\
    \midrule
    Gemini-2.5-Flash \\\midrule
    \hspace{1em}No attack & -- & 57.5 & -- & 57.5 & -- &  57.5  & -- & 74.3 & -- & 61.7\\
    \hspace{1em}No defense & 98.5 & 0.5 & 40.5 & 46.3 & 15 & 57.3 & 53.5 & 64.3 & 51.9 & 42.1 \\\arrayrulecolor{black!30}\midrule
    \hspace{1em}Delimiter & -- & -- & 40.8 & 48.5 & -- & -- & -- & -- &  -- & -- \\
    \hspace{1em}Paraphrasing & 71.8 & 24.0 & -- & -- & -- & -- & 57.3 & 67.3 & -- & -- \\
    \hspace{1em}DP Rewriting & 70.5 & 30.0 & -- & -- & -- & -- & -- &-- & -- & -- \\
    \hspace{1em}GuardRail & 0.0 & 24.5 & 0.0 & 35.3 & 0.0 & 58.5 & 0.0 & 66.3 & 0.0 & 40.2\\
    \hspace{1em}\ours{}& \textbf{0.0} & \textbf{50.5} & \textbf{0.0} & \textbf{55.8} &\textbf{0.0}&\textbf{69.0}&\textbf{0.0}&\textbf{77.7}&\textbf{0.0}&\textbf{63.3} \\\arrayrulecolor{black}\midrule
    Gemini-2.5-Pro \\\midrule
    \hspace{1em}No attack & -- &76.0 & -- & 76.0 & -- & 76.0 & -- & 78.0 & -- & 76.5 \\
    \hspace{1em}No defense & 83.0 & 3.5 &62.3& 68.0& 11.0&79.8&52.2&75.5&52.1&56.7
    \\\arrayrulecolor{black!30}\midrule
    \hspace{1em}GuardRail& \textbf{0.0} & 48.8 & \textbf{0.0}&18.0 &\textbf{0.0}&67.3&\textbf{0.0}&\textbf{72.0}&\textbf{0.0}&51.5\\
    \hspace{1em}\ours{} & \textbf{0.0} & \textbf{55.6}& \textbf{0.0}&\textbf{65.5}&\textbf{0.0}&\textbf{76.8}&\textbf{0.0}&71.3&\textbf{0.0}&\textbf{67.3} 
    \\\arrayrulecolor{black}\midrule
    GPT-4.1 \\\midrule
    \hspace{1em}No attack & -- & 64.5 & -- & 64.5 & -- & 64.5 & -- & 87.0 & -- & 70.1 \\
    \hspace{1em}No defense & 92.5 & 1.0 & 60.0 & 45.3 & 2.8 & 62.3 & 99.5 & 87.0& 63.7& 43.1
    \\\arrayrulecolor{black!30}\midrule
    \hspace{1em}Delimiter & --& -- &64.3&\textbf{52.0}& -- & -- &-- & -- & -- & --\\
    \hspace{1em}Paraphrasing & 80.3& 19.0 & -- &-- &-- & -- & 60.0 & 85.5 & -- & --\\
    \hspace{1em}DP Rewriting  & 74.5 & 15.5 & -- & -- & -- &-- &--&--& -- & --\\
    \hspace{1em}GuardRail& \textbf{0.0} & 20.0 & \textbf{0.0} & 31.5 & \textbf{0.0} & 63.0 & \textbf{0.0} & 82.0& \textbf{0.0} & 44.6 \\
    \hspace{1em}\ours{} & \textbf{0.0} & \textbf{28.0}  & \textbf{0.0} & 42.3 & \textbf{0.0} & \textbf{63.5}& \textbf{0.0} & \textbf{94.5}& \textbf{0.0} &\textbf{57.1} \\\arrayrulecolor{black}\midrule
    Claude-sonnet-4 \\\midrule
    \hspace{1em}No attack & -- & 100.0 & -- &100.0  & -- & 100.0 & -- & 99.0 & -- & 99.8 \\
    \hspace{1em}No defense & 31.3 & 89.0 & 63.8 & 97.0 & 24.0  & 82.0 & 80.5 & 87.8 & 49.9 & 89.0 
    \\\arrayrulecolor{black!30}\midrule
    \hspace{1em}Delimiter & --& -- &60.8 & \textbf{98.3} & -- & -- &-- & -- & -- & -- \\
    \hspace{1em}Paraphrasing & 39.8& \textbf{88.5} & -- &-- &-- & -- & 73.3  & \textbf{90.5} & -- & -- \\
    \hspace{1em}DP Rewriting  & 66.8 & 57.5 & -- & -- & -- &-- &--&--& -- & -- \\
    \hspace{1em}GuardRail& \textbf{0.0}  & 68.5 &\textbf{0.0}& 46.0 & \textbf{0.0}   & 75.5 & \textbf{0.0} & 83.5 & \textbf{0.0} &68.3\\
    \hspace{1em}\ours{}&  \textbf{0.0} & 86.8 & \textbf{0.0} & 91.5 & \textbf{0.0} & \textbf{81.8} &\textbf{0.0}&80.2&\textbf{0.0} & \textbf{85.1}\\
  \arrayrulecolor{black}\bottomrule
\end{tabular}}

\end{table}

\begin{table}[ht]
\resizebox{\textwidth}{!}{%
\begin{threeparttable}
\caption{Performance comparison on the EICU-AC and Mind2Web-SC benchmarks}
\label{table: eicu}

\begin{tabular}{lcccccc}
\toprule
\multicolumn{1}{c}{\textbf{Methods}} & \multicolumn{3}{c}{\textbf{EICU-AC}} & \multicolumn{3}{c}{\textbf{Mind2Web-SC}} \\
\cmidrule(lr){2-4} \cmidrule(lr){5-7}
 & Acc. & P & R & Acc. & P & R \\
\midrule
Hardcoded Safety Rules \citep{xiang2024guardagent}\tnote{*} & 81.0 & 76.6 & 90.7 & 77.5 & 95.1 & 58.0 \\
LLaMA-Guard3 \citep{luo-etal-2025-agrail}\tnote{*} & 48.7 &-- &-- & 56.0 &93.0& 13.0 \\
AgentMonitor (GPT-4o) \citep{luo-etal-2025-agrail}\tnote{*} &82.3 &98.2& 66.7& 72.5& 79.2& 61.0\\
GuardAgent (GPT-4) \citep{xiang2024guardagent}\tnote{*}             & 98.7 & \textbf{100.0}  & 97.5 & 90.0 & \textbf{100.0}  & 80.0 \\
AGrail (GPT-4o) \citep{luo-etal-2025-agrail}\tnote{*}  & 97.8 & 97.5 & 98.1 &\textbf{98.4} &99.0&98.0\\
\ours{} (GPT-4o) & \textbf{100.0} & \textbf{100.0} & \textbf{100.0} & 95.1 & 91.3 & 99.0 \\
\ours{} (GPT-4.1) & \textbf{100.0} & \textbf{100.0} & \textbf{100.0} & 96.2 & 91.2 & \textbf{100.0} \\
\ours{} (Gemini-2.5-pro) & \textbf{100.0} & \textbf{100.0} & \textbf{100.0} & 97.2 & 95.0 & 99.0 \\
\bottomrule
\end{tabular}
\begin{tablenotes}
  \small
  \item[*] Values obtained from the cited papers.
\end{tablenotes}
\end{threeparttable}}
\end{table}

\subsection{Results}

Table~\ref{table: asb} summarizes our evaluation on the ASB dataset, conducted across three backbone LLMs to assess generalization. The table reports the ASR and TSR against several baselines, including a \quotes{No Defense} scenario (providing a lower bound for ASR) and a \quotes{No Attack} scenario (an upper bound for TSR). The low ASR achieved by GuardRail indicates that simple violation detection is a largely solved task for strong LLMs. The primary challenge, therefore, is not if a violation occurs, but how to intervene precisely by blocking only the malicious component (e.g., a specific tool) without degrading task utility. Paraphrasing and Delimiter defenses show high TSR whit Claude-Sonnet-4, Claude-Sonnet showed strong performance in this benchmark \citep{zhangASB}, however the ASR remains high.   \ours{} proves particularly effective at this, achieving a near-zero ASR while simultaneously outperforming all other defenses in TSR, demonstrating a superior trade-off between security and utility.

Table~\ref{table: eicu} summarizes the performance evaluation on the EICU-AC and Mind2Web-SC datasets. To ensure a fair comparison, we use GPT-4o as the backbone LLM, consistent with the SOTA model. We also report with Gemini-2.5-pro. \ours{}, achieves perfect accuracy on the EICU-AC dataset and outperforms all baselines on recall in Mind2Web-SC. This is particularly noteworthy given that \ours{} is a generic policy constructor, whereas a strong baseline like GuardAgent employs a predefined policy structure specifically tailored to these access control tasks. Furthermore, unlike GuardAgent, our method does not require any in-context learning to build its policies. On the other hand, Agrail shows better accuracy and precision showing that an external memory bank of policies can be beneficial. Future, work can enhance \ours{} with memory of previous judgments.
While our method attains high accuracy, we argue that recall is a more critical metric for security applications. On both datasets, \ours{} achieves high recall, signifying that it successfully identifies and blocks every policy violation. This capacity to prevent all illicit actions, even at the cost of a  decrease in precision, is a crucial requirement for deploying secure agentic systems.
\section{Analysis}

\begin{figure}[htbp]
    \centering
    \vspace{-3mm}
    \includegraphics[scale=0.12]
    {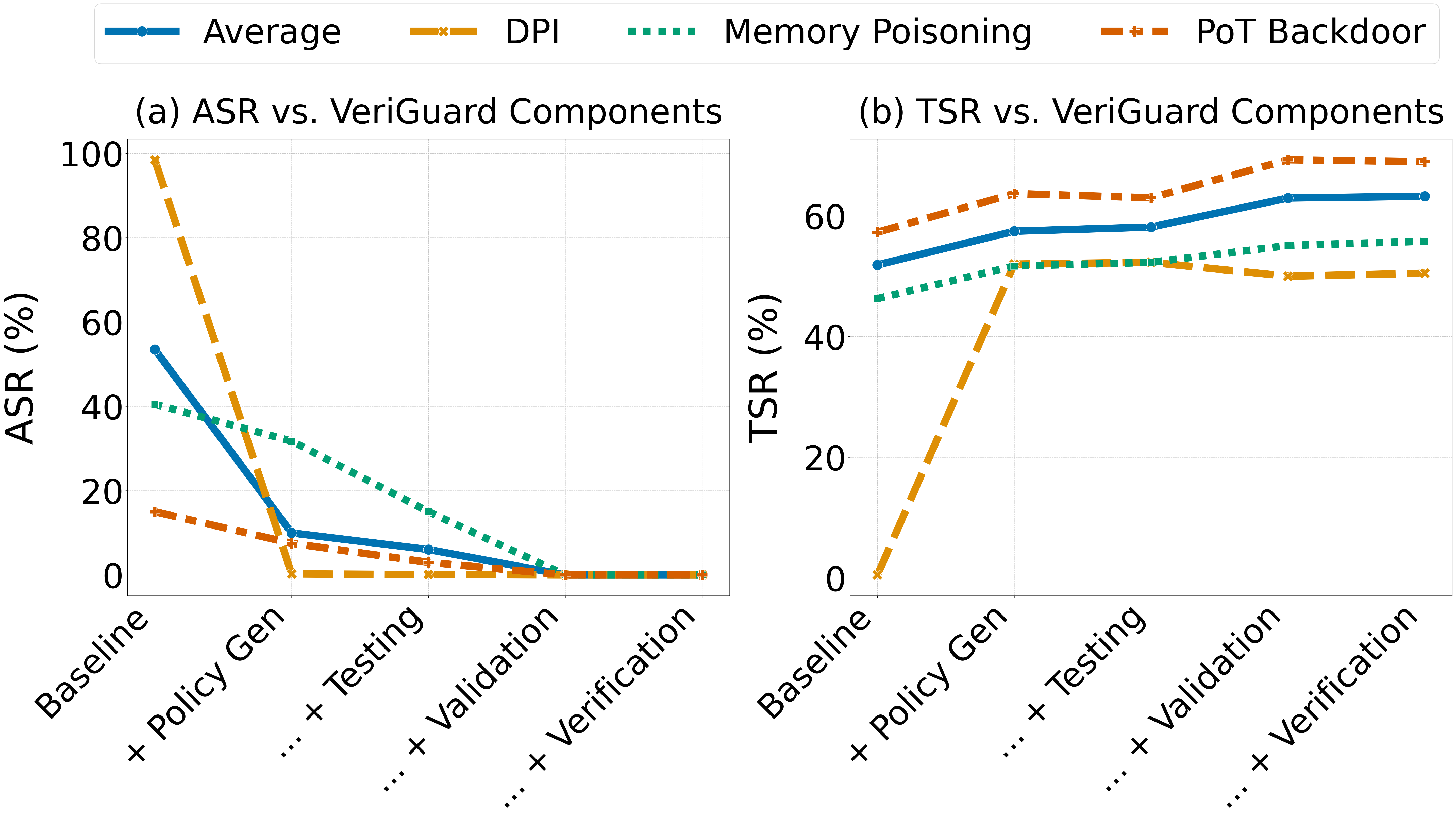}
    \vspace{-5mm}
    \caption{\textbf{(a)} shows the ASR is systematically reduced to 0\% across all evaluated attack types. \textbf{(b)} shows the TSR increases as defense layers are added.} 
    \label{fig:ablation} 
\vspace{-3mm}
\end{figure}

\subsection{Ablation Study of \ours{} Components}

The results of our ablation study, presented in Figure~\ref{fig:ablation}, detail the cumulative impact of each \ours{} component. The analysis was conducted on the Agent Security Benchmark (ASB), utilizing Gemini-2.5-Flash with default parameters. 

Figure~\ref{fig:ablation}a shows  the defense is built in stages as initially the agent is highly vulnerable, with an average ASR of 53.5\%. First, the Policy Generation step provides a substantial impact, reducing the average ASR to 9.97\%. Subsequently, the Validation plays a critical role for complex attacks where the initial policy may be incomplete or non-executable; this is most evident against Memory Poisoning, where this step reduces the ASR by more than half (from 31.75\% to 15\%). Following this, the Validation component further enhances robustness, fully neutralizing all remaining threats and reducing the ASR to 0\% across all attack vectors. Finally, the formal verification step ensures that the defense code rigorously follows all security constraints. Figure~\ref{fig:ablation}b demonstrates that these robust security gains do not incur a performance trade-off. The TSR remains high and exhibits a consistent increase (from 51.87\% to 63.25\% average), confirming \ours{}'s ability to secure the agent without compromising functional utility.

\subsection{Evaluating Integration Methods: Security vs. Utility}

Each strategy offers a different trade-off. Task Termination (TT) is the most stringent approach; it neutralizes threats by terminating any task when an attack is detected. This method is impractical for real-world scenarios because it results in a complete task failure (0\% TSR). Action Blocking (AB) is a less severe strategy that blocks a single malicious action but allows subsequent, non-malicious actions to proceed, forcing the agent to replan. Tool Execution Halt (TEH) offers a more granular approach. A single agent "action" can invoke multiple tool calls (some benign), so TEH blocks only the suspicious tool call—not the entire action—letting the agent continue its plan with a "no tool response" error. In contrast, Collaborative Re-planning (CRP) is the least invasive method. Instead of blocking, \ours{} sends an alert to the agent, which allows it to formulate a new, safer plan. While this significantly boosts the TSR, it doesn't guarantee security, as the agent can still perform unsafe actions (leading to an 11.9\% average ASR). Therefore, a hybrid CRP + TEH approach yields the optimal results. This combination leverages the high TSR of CRP with the fine-grained security of TEH, achieving both a near-zero average ASR (0.1\%) and the highest average TSR (63.6\%).

Table~\ref{tab:integration-results} presents the results from the ASB using Gemini-2.5-Flash. It evaluates the five integration strategies detailed in Section~\ref{subsec:integration}: Task Termination (TT), Action Blocking (AB), Tool Execution Halt (TEH), Collaborative Re-planning (CRP), and a combination of CRP and TEH.

\begin{table}[ht]
\centering
\caption{ASR and TSR for \ours{} integration methods including Task Termination (TT), Action Blocking (AB), Tool Execution Halt (TEH), Collaborative Re-planning (CRP). CRP + TEH combination achieves the optimal balance of security and utility.}
\label{tab:integration-results}
\resizebox{\textwidth}{!}{%
\begin{tabular}{p{2.5cm}ccccccccccc}
\toprule
\multirow{2}{*}{\textbf{Integration Method}} & \multicolumn{2}{c}{\textbf{DPI}} & \multicolumn{2}{c}{\textbf{IPI}} & \multicolumn{2}{c}{\textbf{MP}} & \multicolumn{2}{c}{\textbf{PoT}} & \multicolumn{2}{c}{\textbf{AVG}} \\
\cmidrule(lr){2-3} \cmidrule(lr){4-5} \cmidrule(lr){6-7} \cmidrule(lr){8-9} \cmidrule(lr){10-11}
 & \multicolumn{1}{c}{ASR$\downarrow$} & \multicolumn{1}{c}{TSR$\uparrow$} & \multicolumn{1}{c}{ASR$\downarrow$} & \multicolumn{1}{c}{TSR$\uparrow$} & \multicolumn{1}{c}{ASR$\downarrow$} & \multicolumn{1}{c}{TSR$\uparrow$} & \multicolumn{1}{c}{ASR$\downarrow$} & \multicolumn{1}{c}{TSR$\uparrow$} & \multicolumn{1}{c}{ASR$\downarrow$} & \multicolumn{1}{c}{TSR$\uparrow$} \\
\midrule
TT & 0.0 & 0.0 & 0.0 & 0.0 & 0.0 & 0.0 & 0.0 & 0.0 & 0.0 & 0.0 \\
AB & 0.0 & 0.5 & 0.0 & 34.0 & 0.0 & 55.5 & 0.0 & 62.3 & 0.0 & 38.1 \\
TEH & 0.0 & 0.3 & 0.0 & 48.8 & 0.0 & 61.5 & 0.0 & 68 & 0.0 & 44.6 \\
CRP & 14.3 & 51.5 & 33.3 & 50.0 & 0.0 & 69.0 & 0.0 & 77.7 & 11.9 & 62.1 \\
CRP + TEH & 0.0 & 50.5 & 0.0 & 55.8 & 0.0 & 69.0 & 0.0 & 77.7 & 0.0 & 63.3 \\
\bottomrule

\end{tabular}%
}
\end{table}

\subsection{Limitations}
It is pertinent to acknowledge some limitations of our current approach that define clear avenues for future research. A primary limitation stems from the reliance on a Large Language Model (LLM) to generate formal constraints from natural language—a process that is inherently non-deterministic and susceptible to error. Consequently, the soundness of the formal verification is contingent upon manual validation by the user to ensure the generated constraints accurately reflect their intent.
Secondly, the system's capabilities are intrinsically bound by the underlying program verification tool, Nagini. As Nagini is an active research project, it may possess a limited grammar for expressing certain complex properties. Furthermore, while extending the framework to other programming languages is possible, doing so represents a non-trivial implementation challenge.
Finally, our hybrid architecture, which integrates an LLM for argument interpretation with deterministic Python code for rule implementation, may be insufficient for identifying sophisticated attacks that require deeper capacity for logical reasoning and dynamic policy updates, which presents a key direction for future investigation.
\section{Conclusion}
\label{sec: conclusion}
In this work, we introduce \ours{}, a novel framework designed to substantially enhance the safety and reliability of Large Language Model (LLM) agents. By integrating a verification module that formally checks agent-generated policies and actions against predefined safety specifications, \ours{} moves beyond reactive, pattern-matching safety measures to a proactive, provably-sound approach. Our experiments demonstrate that this interactive verification loop is highly effective at preventing a wide range of unsafe operations, from prompt injections to unauthorized data access, while maintaining a high degree of task success. The results on benchmarks such as ASB, EICU-AC, and Mind2Web-SC show that VeriGuard not only significantly reduces the attack success rate to near-zero but also offers flexible policy enforcement strategies that can be tailored to different operational needs. \ours{} provides a robust and essential safeguard, paving the way for the trustworthy deployment of LLM agents in complex and high-stakes real-world environments.

Building on the foundation of \ours{}, several promising avenues for future research emerge. One key direction is the scalability and efficiency of the formal verification process. Another area for exploration is the autonomous generation of safety specifications themselves.

\bibliographystyle{abbrvnat}
\nobibliography*
\bibliography{main}

\begin{thebibliography}{33}
\providecommand{\natexlab}[1]{#1}
\providecommand{\url}[1]{\texttt{#1}}
\expandafter\ifx\csname urlstyle\endcsname\relax
  \providecommand{\doi}[1]{doi: #1}\else
  \providecommand{\doi}{doi: \begingroup \urlstyle{rm}\Url}\fi

\bibitem[Bai et~al.(2022)Bai, Kadavath, Kundu, Askell, Kernion, Jones, Chen, Goldie, Mirhoseini, McKinnon, Chen, Olsson, Olah, Hernandez, Drain, Ganguli, Li, Tran-Johnson, Perez, Kerr, Mueller, Ladish, Landau, Ndousse, Lovitt, Sellitto, Elhage, Schiefer, Mercado, DasSarma, Lasenby, Grosse, Ringer, Johnston, Kravec, Showk, Fort, Lanham, Telleen-Lawton, Conerly, Henighan, Hume, Bowman, Hatfield-Dodds, Mann, Amodei, Joseph, McCandlish, Brown, and Kaplan]{bai2022constitutional}
Y.~Bai, S.~Kadavath, S.~Kundu, A.~Askell, J.~Kernion, A.~Jones, A.~Chen, A.~Goldie, A.~Mirhoseini, C.~McKinnon, C.~Chen, C.~Olsson, C.~Olah, D.~Hernandez, D.~Drain, D.~Ganguli, D.~Li, E.~Tran-Johnson, E.~Perez, J.~Kerr, J.~Mueller, J.~Ladish, J.~Landau, K.~Ndousse, L.~Lovitt, M.~Sellitto, N.~Elhage, N.~Schiefer, N.~Mercado, N.~DasSarma, R.~Lasenby, R.~Grosse, S.~Ringer, S.~Johnston, S.~Kravec, S.~E. Showk, S.~Fort, T.~Lanham, T.~Telleen-Lawton, T.~Conerly, T.~Henighan, T.~Hume, S.~R. Bowman, Z.~Hatfield-Dodds, B.~Mann, D.~Amodei, N.~Joseph, S.~McCandlish, T.~Brown, and J.~Kaplan.
\newblock Constitutional ai: Harmlessness from ai feedback, 2022.

\bibitem[Chan et~al.(2024)Chan, Yu, Chen, Jiang, Liu, Shi, Liu, Xue, and Guo]{chan2024agentmonitorplugandplayframeworkpredictive}
C.-M. Chan, J.~Yu, W.~Chen, C.~Jiang, X.~Liu, W.~Shi, Z.~Liu, W.~Xue, and Y.~Guo.
\newblock Agentmonitor: A plug-and-play framework for predictive and secure multi-agent systems, 2024.
\newblock URL \url{https://arxiv.org/abs/2408.14972}.

\bibitem[Chen et~al.(2025)Chen, Kang, and Li]{chen2025shieldagent}
Z.~Chen, M.~Kang, and B.~Li.
\newblock Shieldagent: Shielding agents via verifiable safety policy reasoning.
\newblock \emph{ICML}, 2025.

\bibitem[Eilers and M{\"u}ller(2018)]{eilers2018nagini}
M.~Eilers and P.~M{\"u}ller.
\newblock Nagini: a static verifier for python.
\newblock In \emph{International Conference on Computer Aided Verification}, pages 596--603. Springer, 2018.

\bibitem[Eilers et~al.(2025)Eilers, Schwerhoff, Summers, and M\"{u}ller]{EilersSchwerhoffSummersMueller25}
M.~Eilers, M.~Schwerhoff, A.~J. Summers, and P.~M\"{u}ller.
\newblock Fifteen years of viper.
\newblock In R.~Piskac and Z.~Rakamari{\'{c}}, editors, \emph{Computer Aided Verification (CAV)}, pages 107--123, Cham, 2025. Springer Nature Switzerland.
\newblock \doi{10.1007/978-3-031-98668-0_5}.
\newblock URL \url{https://link.springer.com/chapter/10.1007/978-3-031-98668-0_5}.

\bibitem[Ganguli et~al.(2022)Ganguli, Lovitt, Kernion, Askell, Bai, Kadavath, Mann, Perez, Schiefer, Ndousse, Jones, Bowman, Chen, Conerly, DasSarma, Drain, Elhage, El-Showk, Fort, Hatfield-Dodds, Henighan, Hernandez, Hume, Jacobson, Johnston, Kravec, Olsson, Ringer, Tran-Johnson, Amodei, Brown, Joseph, McCandlish, Olah, Kaplan, and Clark]{perez2022red}
D.~Ganguli, L.~Lovitt, J.~Kernion, A.~Askell, Y.~Bai, S.~Kadavath, B.~Mann, E.~Perez, N.~Schiefer, K.~Ndousse, A.~Jones, S.~Bowman, A.~Chen, T.~Conerly, N.~DasSarma, D.~Drain, N.~Elhage, S.~El-Showk, S.~Fort, Z.~Hatfield-Dodds, T.~Henighan, D.~Hernandez, T.~Hume, J.~Jacobson, S.~Johnston, S.~Kravec, C.~Olsson, S.~Ringer, E.~Tran-Johnson, D.~Amodei, T.~Brown, N.~Joseph, S.~McCandlish, C.~Olah, J.~Kaplan, and J.~Clark.
\newblock Red teaming language models to reduce harms: Methods, scaling behaviors, and lessons learned, 2022.
\newblock URL \url{https://arxiv.org/abs/2209.07858}.

\bibitem[Ghaffarian et~al.(2024)Ghaffarian, Raval, Bavota, and Izadi]{ghaffarian2024can}
S.~Ghaffarian, R.~Raval, G.~Bavota, and M.~Izadi.
\newblock Can llms verify their own code? a case study in secure web development, 2024.

\bibitem[Gou et~al.(2025)Gou, Huang, Ning, Gu, Lin, Qi, Kopanev, Yu, Gutiérrez, Shu, Song, Wu, Chen, Moussa, Zhang, Xie, Li, Xue, Liao, Zhang, Zheng, Cai, Rozgic, Ziyadi, Sun, and Su]{gou2025mind2web2}
B.~Gou, Z.~Huang, Y.~Ning, Y.~Gu, M.~Lin, W.~Qi, A.~Kopanev, B.~Yu, B.~J. Gutiérrez, Y.~Shu, C.~H. Song, J.~Wu, S.~Chen, H.~N. Moussa, T.~Zhang, J.~Xie, Y.~Li, T.~Xue, Z.~Liao, K.~Zhang, B.~Zheng, Z.~Cai, V.~Rozgic, M.~Ziyadi, H.~Sun, and Y.~Su.
\newblock Mind2web 2: Evaluating agentic search with agent-as-a-judge, 2025.

\bibitem[Gravitas(2023)]{AutoGPT}
S.~Gravitas.
\newblock Auto-gpt: An autonomous gpt-4 experiment.
\newblock \url{https://github.com/Significant-Gravitas/Auto-GPT}, 2023.

\bibitem[Inan et~al.(2023)Inan, Kandasamy, Rameshbabu, El-Khamy, Purohit, and Ranganath]{meta2024llama_guard}
H.~Inan, K.~Kandasamy, S.~Rameshbabu, M.~El-Khamy, S.~Purohit, and S.~Ranganath.
\newblock Llama guard: Llm-based input-output safeguard for human-ai conversations, 2023.
\newblock URL \url{https://ai.meta.com/research/publications/llama-guard-llm-based-input-output-safeguard-for-human-ai-conversations/}.

\bibitem[Jain et~al.(2023)Jain, Schwarzschild, Wen, Somepalli, Kirchenbauer, yeh Chiang, Goldblum, Saha, Geiping, and Goldstein]{jain2023baselinedefensesadversarialattacks}
N.~Jain, A.~Schwarzschild, Y.~Wen, G.~Somepalli, J.~Kirchenbauer, P.~yeh Chiang, M.~Goldblum, A.~Saha, J.~Geiping, and T.~Goldstein.
\newblock Baseline defenses for adversarial attacks against aligned language models, 2023.
\newblock URL \url{https://arxiv.org/abs/2309.00614}.

\bibitem[Jin et~al.(2024)Jin, Zhang, Zhou, Li, Gao, and Chen]{jin2024llm}
Z.~Jin, H.~Zhang, Z.~Zhou, J.~Li, M.~Gao, and E.~Chen.
\newblock Llm-safeguard: A human-in-the-loop framework for tuning safety-guard of llm-based agents, 2024.

\bibitem[Li et~al.(2024)Li, Zhang, Chen, Wang, Wang, Chen, Li, Tang, ud~K.~Effendy, Nguyen, Xie, Tsai, and Chen]{li2024llmbased}
G.~Li, Y.~Zhang, Z.~Chen, H.~Wang, Z.~Wang, S.-Q. Chen, Y.-F. Li, Z.~Tang, M.~ud~K.~Effendy, A.-T.~T. Nguyen, X.~Xie, M.-H. Tsai, and T.-C. Chen.
\newblock Llm-based generation of verifiable computation, 2024.

\bibitem[Llama~Team(2024)]{dubey2024llama3herdmodels}
A.~.~M. Llama~Team.
\newblock The llama 3 herd of models, 2024.
\newblock URL \url{https://arxiv.org/abs/2407.21783}.

\bibitem[Luo et~al.(2025)Luo, Dai, Liu, Banerjee, Sun, Chen, and Xiao]{luo-etal-2025-agrail}
W.~Luo, S.~Dai, X.~Liu, S.~Banerjee, H.~Sun, M.~Chen, and C.~Xiao.
\newblock {AG}rail: A lifelong agent guardrail with effective and adaptive safety detection.
\newblock In W.~Che, J.~Nabende, E.~Shutova, and M.~T. Pilehvar, editors, \emph{Proceedings of the 63rd Annual Meeting of the Association for Computational Linguistics (Volume 1: Long Papers)}, pages 8104--8139, Vienna, Austria, July 2025. Association for Computational Linguistics.
\newblock ISBN 979-8-89176-251-0.
\newblock \doi{10.18653/v1/2025.acl-long.399}.
\newblock URL \url{https://aclanthology.org/2025.acl-long.399/}.

\bibitem[Mattern et~al.(2023)Mattern, Mireshghallah, Jin, Schoelkopf, Sachan, and Berg-Kirkpatrick]{mattern-etal-2023-membership}
J.~Mattern, F.~Mireshghallah, Z.~Jin, B.~Schoelkopf, M.~Sachan, and T.~Berg-Kirkpatrick.
\newblock Membership inference attacks against language models via neighbourhood comparison.
\newblock In A.~Rogers, J.~Boyd-Graber, and N.~Okazaki, editors, \emph{Findings of the Association for Computational Linguistics: ACL 2023}, pages 11330--11343, Toronto, Canada, July 2023. Association for Computational Linguistics.
\newblock \doi{10.18653/v1/2023.findings-acl.719}.
\newblock URL \url{https://aclanthology.org/2023.findings-acl.719/}.

\bibitem[Pan et~al.(2024)Pan, Saxon, Xu, Nathani, Wang, and Wang]{pan-etal-2024-automatically}
L.~Pan, M.~Saxon, W.~Xu, D.~Nathani, X.~Wang, and W.~Y. Wang.
\newblock Automatically correcting large language models: Surveying the landscape of diverse automated correction strategies.
\newblock \emph{Transactions of the Association for Computational Linguistics}, 12:\penalty0 484--506, 2024.
\newblock \doi{10.1162/tacl_a_00660}.
\newblock URL \url{https://aclanthology.org/2024.tacl-1.27/}.

\bibitem[Park et~al.(2023)Park, O'Brien, Cai, Morris, Liang, and Bernstein]{park2023generative}
J.~S. Park, J.~C. O'Brien, C.~J. Cai, M.~R. Morris, P.~Liang, and M.~S. Bernstein.
\newblock Generative agents: Interactive simulacra of human behavior, 2023.

\bibitem[Patil et~al.(2024)Patil, Zhang, Wang, and Gonzalez]{patil2023gorilla}
S.~G. Patil, T.~Zhang, X.~Wang, and J.~E. Gonzalez.
\newblock Gorilla: Large language model connected with massive apis.
\newblock In A.~Globerson, L.~Mackey, D.~Belgrave, A.~Fan, U.~Paquet, J.~Tomczak, and C.~Zhang, editors, \emph{Advances in Neural Information Processing Systems}, volume~37, pages 126544--126565. Curran Associates, Inc., 2024.
\newblock URL \url{https://proceedings.neurips.cc/paper_files/paper/2024/file/e4c61f578ff07830f5c37378dd3ecb0d-Paper-Conference.pdf}.

\bibitem[Rebedea et~al.(2023)Rebedea, Dinu, Sreedhar, Parisien, and Cohen]{rebedea-etal-2023-nemo}
T.~Rebedea, R.~Dinu, M.~N. Sreedhar, C.~Parisien, and J.~Cohen.
\newblock {N}e{M}o guardrails: A toolkit for controllable and safe {LLM} applications with programmable rails.
\newblock In Y.~Feng and E.~Lefever, editors, \emph{Proceedings of the 2023 Conference on Empirical Methods in Natural Language Processing: System Demonstrations}, pages 431--445, Singapore, Dec. 2023. Association for Computational Linguistics.
\newblock \doi{10.18653/v1/2023.emnlp-demo.40}.
\newblock URL \url{https://aclanthology.org/2023.emnlp-demo.40}.

\bibitem[Schick et~al.(2023{\natexlab{a}})Schick, Dwivedi-Yu, Dess{\`\i}, Raileanu, Tsvigun, Cancedda, and Scialom]{schick2023toolformer}
T.~Schick, J.~Dwivedi-Yu, R.~Dess{\`\i}, R.~Raileanu, M.~Tsvigun, N.~Cancedda, and T.~Scialom.
\newblock Toolformer: Language models can teach themselves to use tools.
\newblock \emph{arXiv preprint arXiv:2302.04761}, 2023{\natexlab{a}}.

\bibitem[Schick et~al.(2023{\natexlab{b}})Schick, Dwivedi-Yu, Dessì, Raileanu, Lomeli, Zettlemoyer, Cancedda, and Scialom]{schick2024toolformer}
T.~Schick, J.~Dwivedi-Yu, R.~Dessì, R.~Raileanu, M.~Lomeli, L.~Zettlemoyer, N.~Cancedda, and T.~Scialom.
\newblock Toolformer: Language models can teach themselves to use tools, 2023{\natexlab{b}}.

\bibitem[Schulhoff et~al.(2023)Schulhoff, Pinto, Khan, Bouchard, Si, Boyd-Graber, Anati, Tagliabue, Kost, and Carnahan]{Schulhoff2022ignore}
S.~V. Schulhoff, J.~Pinto, A.~Khan, L.-F. Bouchard, C.~Si, J.~L. Boyd-Graber, S.~Anati, V.~Tagliabue, A.~L. Kost, and C.~R. Carnahan.
\newblock Ignore this title and hackaprompt: Exposing systemic vulnerabilities of llms through a global prompt hacking competition.
\newblock In \emph{Empirical Methods in Natural Language Processing}, 2023.

\bibitem[Shinn et~al.(2023)Shinn, Cassano, Gopinath, Narasimhan, and Yao]{shinn2023reflexion}
N.~Shinn, F.~Cassano, A.~Gopinath, K.~Narasimhan, and S.~Yao.
\newblock Reflexion: Language agents with verbal reinforcement learning, 2023.

\bibitem[Wang et~al.(2023)Wang, Ma, Feng, Zhang, Yang, Zhang, Chen, Tang, Chen, Lin, Zhao, Wei, and Wen]{wang2023survey}
L.~Wang, C.~Ma, X.~Feng, Z.~Zhang, H.~Yang, J.~Zhang, Z.-Y. Chen, J.~Tang, X.~Chen, Y.~Lin, W.~X. Zhao, Z.~Wei, and J.-R. Wen.
\newblock A survey on large language model based autonomous agents, 2023.

\bibitem[Wei et~al.(2023)Wei, Haghtalab, and Steinhardt]{wei2023jailbroken}
A.~Wei, N.~Haghtalab, and J.~Steinhardt.
\newblock Jailbroken: How does llm safety training fail?
\newblock \emph{NeurIPS}, 2023.

\bibitem[Xi et~al.(2023)Xi, Chen, Guo, He, Ding, Hong, Zhang, Wang, Jin, Zhou, et~al.]{xi2023rise}
Z.~Xi, W.~Chen, X.~Guo, W.~He, Y.~Ding, B.~Hong, M.~Zhang, J.~Wang, S.~Jin, E.~Zhou, et~al.
\newblock The rise and potential of large language model based agents: A survey.
\newblock \emph{arXiv preprint arXiv:2309.07864}, 2023.

\bibitem[Xiang et~al.(2025)Xiang, Zheng, Li, Hong, Li, Xie, Zhang, Xiong, Xie, Yang, Song, and Li]{xiang2024guardagent}
Z.~Xiang, L.~Zheng, Y.~Li, J.~Hong, Q.~Li, H.~Xie, J.~Zhang, Z.~Xiong, C.~Xie, C.~Yang, D.~Song, and B.~Li.
\newblock Guardagent: Safeguard llm agents by a guard agent via knowledge-enabled reasoning, 2025.

\bibitem[Xu et~al.(2023)Xu, Kong, Liu, Cui, Wang, Zhang, and Kankanhalli]{xu2023llmfoolitselfpromptbased}
X.~Xu, K.~Kong, N.~Liu, L.~Cui, D.~Wang, J.~Zhang, and M.~Kankanhalli.
\newblock An llm can fool itself: A prompt-based adversarial attack, 2023.
\newblock URL \url{https://arxiv.org/abs/2310.13345}.

\bibitem[Yao et~al.(2023)Yao, Zhao, Yu, Du, Shafran, Narasimhan, and Cao]{yao2022react}
S.~Yao, J.~Zhao, D.~Yu, N.~Du, I.~Shafran, K.~Narasimhan, and Y.~Cao.
\newblock React: Synergizing reasoning and acting in language models, 2023.

\bibitem[Zhang et~al.(2025)Zhang, Huang, Mei, Yao, Wang, Zhan, Wang, and Zhang]{zhangASB}
H.~Zhang, J.~Huang, K.~Mei, Y.~Yao, Z.~Wang, C.~Zhan, H.~Wang, and Y.~Zhang.
\newblock Agent security bench (asb): Formalizing and benchmarking attacks and defenses in llm-based agents.
\newblock In \emph{The Thirteenth International Conference on Learning Representations}, 2025.

\bibitem[Zhao et~al.(2025)Zhao, Chen, Zhang, and Li]{zhao2025recodeimprovingllmbasedcode}
Y.~Zhao, S.~Chen, J.~Zhang, and Z.~Li.
\newblock Recode: Improving llm-based code repair with fine-grained retrieval-augmented generation, 2025.
\newblock URL \url{https://arxiv.org/abs/2509.02330}.

\bibitem[Zhou et~al.(2023)Zhou, Xu, Zhu, Zhou, Lo, Sridhar, Cheng, Ou, Bisk, Fried, et~al.]{zhou2023webarena}
S.~Zhou, F.~F. Xu, H.~Zhu, X.~Zhou, R.~Lo, A.~Sridhar, X.~Cheng, T.~Ou, Y.~Bisk, D.~Fried, et~al.
\newblock Webarena: A realistic web environment for building autonomous agents.
\newblock \emph{arXiv preprint arXiv:2307.13854}, 2023.

\end{thebibliography}

\clearpage
\appendix
\section{Experiment Details}~\label{app: exp_detail}
This section details the implementation VeriGuard with agent systems mentioned in Section~\ref{sec: exp}.

\subsection{Prompts Used To Generate the Policy}\label{ap:prompt_generation}

\begin{tcolorbox}[
  title=Prompt: Policy Code Generation,
  colback=PromptBackground,
  colframe=PromptFrame,
  fonttitle=\bfseries\color{PromptTitle},
  breakable, 
  arc=2mm,   
]

You are an expert AI security agent. Your primary function is to generate a Python security policy function based on a high-level user request.

\newcommand{\promptsection}[1]{\par\medskip\textcolor{PromptSection}{\bfseries \#\# #1}\par\medskip}
\newcommand{\promptsubsection}[1]{\par\medskip\textcolor{PromptSection!80!black}{\bfseries \#\#\# #1}\par\medskip}

\promptsection{CONTEXT}

You are part of a system that moderates a target agent's actions at runtime. This system works as follows:

1. A user gives a natural language security request (e.g., ``The agent cannot send emails to non-company addresses").

2. You receive this request and generate a specific Python \var{policy} function to enforce it.

3. At runtime, before the target agent acts, its inputs and outputs are intercepted.

4. Parameters are extracted from this runtime data using the \var{read\_parameters\_with\_LLM} tool.

5. The extracted parameters are passed to your generated \var{policy} function, which decides whether to allow or deny the action.

**Runtime Data:**

- \var{agent\_information: str} \textcolor{PromptComment}{\# The text prompt input given to the agent, the agent output, and any available context.}

**Parameter Extraction Tool:**
The system uses the \var{read\_parameters\_with\_LLM} tool to extract relevant arguments from the \var{agent\_information}.

- **Tool Signature**: \var{read\_parameters\_with\_LLM(text: str, list\_of\_arguments\_to\_read: dict[str, type]) -$>$ dict}

- **Example**: Given the text \textit{``Send an email to alex@example.com and Dona Jhonson with the subject `Update'"} and \var{list\_of\_arguments\_to\_read=\{\{``recipients": list, ``subject": str\}\}}, the tool would return \var{\{\{``recipients": [``alex@example.com", ``dona@example.com"], ``subject": ``Update"\}\}}.

\promptsection{YOUR TASK}

Given a agent's security \var{request} and the target \var{agent\_specification}, you must generate two outputs: a list of parameters to extract, and the Python code for the policy function.

\promptsection{OUTPUT REQUIREMENTS}

Your response must contain exactly two parts in the following order:

**1. Arguments to Extract** \par
A Python dictionary named \var{list\_of\_arguments\_to\_read} that contains all the string arguments needed from the runtime data to evaluate the policy.
Follow this exact format:
\begin{lstlisting}[style=mypython]
```python
 list_of_arguments_to_read = {{"argument1": type1, "argument2": type2, "argument3": type3}}
```
\end{lstlisting}

**2. Policy Function Header**
\begin{lstlisting}[style=mypython]
```python
def policy(arguments: dict) -> tuple[bool, str]:
```
\end{lstlisting}

**3. Constant variables** \par
You can define global variables needed to execute the function.
\begin{lstlisting}[style=mypython]
```python
variable1 = "value1"
variable2 = ["value2"]
```
\end{lstlisting}

**4. Policy Function Code** \par
A Python function that implements the user's security policy. You can create auxiliary functions as well, but the policy function must be named exactly \var{policy}.

The function must adhere to these rules:

- It must be named exactly \var{policy}.

- It must accept one argument: \var{arguments: dict}.

- It must return a tuple: \var{(bool, str)}. The \var{bool} is the decision: \var{True} if the action is **allowed**, \var{False} if it is **denied**.

- It must include a clear docstring explaining the policy.

- It must not use any external libraries.

Note: Do not use python string functions to get parameters or arguments. Use the \var{read\_parameters\_with\_LLM} tool instead.

Follow this exact format:
\begin{lstlisting}[style=mypython]
```python
def policy(arguments: dict) -> tuple[bool, str]:
    """A brief, one-sentence description of what this policy enforces."""
    # Your policy logic here
    # ...
    if condition_is_met:
        return True, "Reason for allowing."
    else:
        return False, "Reason for denial."
```
\end{lstlisting}

---

Target Agent Specification:

\inputvar{\{agent\_specification\}}

User Request

\inputvar{\{request\}}

Requirements \textcolor{PromptComment}{\# If available}

\inputvar{\{requirements\}}

Assumptions \textcolor{PromptComment}{\# If available}

\inputvar{\{assumptions\}}

\end{tcolorbox}

\begin{tcolorbox}[
  title=Prompt: Constraint Generation,
  colback=PromptBackground,
  colframe=PromptFrame,
  fonttitle=\bfseries\color{PromptTitle},
  breakable, 
  arc=2mm,   
]
\label{prompt:pg}
\newcommand{\promptsection}[1]{\par\medskip\textcolor{PromptSection}{\bfseries \#\# #1}\par\medskip}
\newcommand{\promptsubsection}[1]{\par\medskip\textcolor{PromptSection!80!black}{\bfseries \#\#\# #1}\par\medskip}

You are a senior **Security Systems Architect** with a deep understanding of **Natural Language Understanding (NLU)** and formal logic. Your expertise lies in translating complex, human-written security policies into clear, unambiguous, and verifiable logical rules.

\promptsection{Objective} 

Your task is to analyze a given natural language security policy, an agent's description, and a set of defined variables/constants to generate a set of discrete, testable logical statements. These statements should represent the core requirements of the policy and can be used to programmatically verify the agent's behavior.

\promptsection{Instructions}

1.  **Analyze the Inputs:** Carefully read the provided security policy, the agent's description, and the list of variables and constants. Identify key actors, actions, conditions, and the specific capabilities of the agent that are relevant to the policy.

2.  **Generate Logical Statements:** For each policy rule, create one or more logical statements.
    * Each statement should be a complete, verifiable condition.
    * Use a clear **IF-THEN** or similar logical structure.
    * Focus on verifiable conditions (e.g., "The agent's state is 'idle'") and outcomes (e.g., "The agent must not respond to external commands").
    * Ensure the statements cover both "positive" (allowed) and "negative" (denied) cases, taking into account the agent's specific functions.
    
3.  **Provide a Rationale:** After the list, briefly explain the logical structure and why these statements effectively capture the policy's intent and apply to the described agent.

4.  **Output Format:**

    Verifiable Logical Statements:
    
    (Give the statements in json format as:)

    \begin{lstlisting}[style=mypython]
    ```json
    [{{"statement": ...}},
    {{"statement": ...}}
    ...
    ]```\end{lstlisting}

Rationale:
    
(A brief explanation of the logical framework and how it aligns with the policy, the agent's capabilities, and the provided variables.)

\promptsection{Input:}
Security Policy Specification:

\inputvar{\{security policy\}}

Agent Specification:

\inputvar{\{agents\_description\}}

Variables and Constants: 

\inputvar{\{variables\_and\_constants\}}

Requirements \textcolor{PromptComment}{\# If available}

\inputvar{\{requirements\}}

Assumptions \textcolor{PromptComment}{\# If available}

\inputvar{\{assumptions\}}

\promptsection{Output:}
Verifiable Logical Statements:
\end{tcolorbox}

\subsection{Prompts Used To Enforce the Policy}\label{ap:prompt_enforce}
\begin{tcolorbox}[
  title=Prompt: Argument Extraction,
  colback=PromptBackground,
  colframe=PromptFrame,
  fonttitle=\bfseries\color{PromptTitle},
  breakable, 
  arc=2mm,   
]

\newcommand{\promptsection}[1]{\par\medskip\textcolor{PromptSection}{\bfseries \#\# #1}\par\medskip}
\newcommand{\promptsubsection}[1]{\par\medskip\textcolor{PromptSection!80!black}{\bfseries \#\#\# #1}\par\medskip}
You are an expert Natural Language Understanding (NLU) engine. Your task is to extract structured information from a given text based on a specification and return it as a single Python dictionary.

\promptsection{YOUR TASK}

You will be provided with a block of input text, a Python function for context, and a specification of the parameters to extract. You must carefully read the **entire** input text, find the values for each required parameter, and format them into a dictionary.

**Instructions:**

  - Synthesize information from all parts of the input text to fill the parameters.
  
  - Ensure the data types in the output dictionary match the specification (e.g., \var{list} for \var{list[str]}).
  
  - Your output at the end should include your reasoning and a valid Python dictionary.

-----

\promptsection{EXAMPLE}

GIVEN THIS CONTEXT:

Python Function:

\begin{lstlisting}[style=mypython]
```python
def send_email(arguments: dict) -> str:
    """Send an email to the specified recipients with the given subject."""
    # send email logic here
    # ...
    return "Email sent successfully."
```
\end{lstlisting}

Agent Information:

\textit{Input: Send an email to alex@example.com and Dona Jhonson with the subject 'Update'}

\textit{Output: Step 1: Using retreive\_email("Dona Jhonson") we get the email "dona@example.com"}

\vspace{5pt}
AND THIS PARAMETER SPECIFICATION:

\var{\{\{"recipients": "list[str]", "subject": "str"\}\}}

\vspace{5pt}
PRODUCE THIS OUTPUT:

**Reasoning**

From the "Agent Input", I found one recipient, "alex@example.com", and the subject, "Updates".

From "Agent Output" Step 1, I found the email for "Dona Jhonson" is "dng@example.com". I will combine these into the `recipients` list.

**Python dictionary**
\begin{lstlisting}[style=mypython]
```python
{"recipients": ["alex@example.com", "dona@example.com"], "subject": "Update"}
```
\end{lstlisting}

-----

\promptsection{YOUR TURN}
\vspace{5pt}
GIVEN THIS CONTEXT:

Python Function:

\inputvar{\{function\}}

Input Text:

\inputvar{\{text\}}

\vspace{5pt}
AND THIS PARAMETER SPECIFICATION:

\inputvar{\{parameters\}}

\vspace{5pt}
PRODUCE THIS OUTPUT:
\end{tcolorbox}

\subsection{Prompts Used for Validation}\label{ap:prompt_validation}
\begin{tcolorbox}[
  title=Prompt: Validation Analysis,
  colback=PromptBackground,
  colframe=PromptFrame,
  fonttitle=\bfseries\color{PromptTitle},
  breakable, 
  arc=2mm,   
]

\newcommand{\promptsection}[1]{\par\medskip\textcolor{PromptSection}{\bfseries \#\# #1}\par\medskip}
\newcommand{\promptsubsection}[1]{\par\medskip\textcolor{PromptSection!80!black}{\bfseries \#\#\# #1}\par\medskip}
You are an expert **Natural Language Understanding (NLU)** and **logic engine**. Your primary function to verify logical statements.

\promptsection{YOUR TASK}
Given a user's security specification and statements, you must analyze the specification in detail and then check if the logical statements is valid or needs correction.

1. Check if the user specification has ambiguity, needs clarification, for example co-references.

2. Check pre-assumptions for the statements. Focus on the specification.

4. Find contra examples.

5. Find any logical error in the statements.

Output:
After your analysis list all the points that require clarification or correction. 

---

User Specification

\inputvar{\{user\_specification\}}

Logical Statements

\inputvar{\{statements\}}
\end{tcolorbox}

\begin{tcolorbox}[
  title=Prompt: Validation Disambiguation,
  colback=PromptBackground,
  colframe=PromptFrame,
  fonttitle=\bfseries\color{PromptTitle},
  breakable, 
  arc=2mm,   
]
\newcommand{\promptsection}[1]{\par\medskip\textcolor{PromptSection}{\bfseries \#\# #1}\par\medskip}
\newcommand{\promptsubsection}[1]{\par\medskip\textcolor{PromptSection!80!black}{\bfseries \#\#\# #1}\par\medskip}
You are an expert in **System Requirements**, **Security Policy**, and **Logical Deduction**. Your primary function is to act as an arbiter to resolve ambiguities identified in a system analysis. You must review a user's security goals, the agent's capabilities, and the provided analysis to establish a definitive, clear, and reasonable set of system requirements and assumptions.

\promptsection{YOUR TASK}
You are given a high-level \var{user\_specification}, the technical \var{agent\_specification}, and an \var{analysis} that identifies points of ambiguity, conflict, or missing details.

Your task is to:

1.  Carefully examine each point raised in the \var{analysis}.

2.  Use the \var{user\_specification} as the primary source of intent and the \var{agent\_specification} as the context for technical constraints.

3.  For each point of ambiguity, make a clear and logical **decision** to finalize the requirement or assumption.

4.  Compile these decisions, along with any original unambiguous requirements, into a single, comprehensive list of detailed requirements.

\promptsection{OUTPUT FORMAT}
Your response must contain two parts:

**Part 1: Decisions on Ambiguities**

For each point from the analysis, provide your decision in the following structured format:

1. [Title of the Point/Ambiguity]

- Decision: [State your clear and final decision on the requirement or assumption.]

- Justification: [Briefly explain *why* this decision is the most reasonable, referencing the user/agent specifications as needed.]

2. [Title of the Next Point/Ambiguity]

- Decision: [...]

- Justification: [...]

**Part 2: Finalized Detailed Requirements List**

After addressing all ambiguities, compile a complete and final list of all detailed requirements (combining the original, clear requirements with your new decisions).

1.  [Detailed Requirement 1]

2.  [Detailed Requirement 2]

3.  [Detailed Requirement 3]

---
\promptsection{INPUTS}

User Specification

\inputvar{\{user\_specification\}}

Agent Specification

\inputvar{\{agent\_specification\}}

Analysis of Ambiguities

\inputvar{\{analysis\}}

\end{tcolorbox}

\subsection{Prompts Used for Code Testing}\label{ap:prompt_testing}
\begin{tcolorbox}[
  title=Prompt: Test Case Generation,
  colback=PromptBackground,
  colframe=PromptFrame,
  fonttitle=\bfseries\color{PromptTitle},
  breakable, 
  arc=2mm,   
]
\newcommand{\promptsection}[1]{\par\medskip\textcolor{PromptSection}{\bfseries \#\# #1}\par\medskip}
\newcommand{\promptsubsection}[1]{\par\medskip\textcolor{PromptSection!80!black}{\bfseries \#\#\# #1}\par\medskip}
You are an expert at writing Pytest functions. Your task is to generate complete and effective test cases for a given Python function, adhering to best practices.

\promptsection{YOUR TASK}
Generate Pytest functions within a single Python code block. The tests should be comprehensive, covering a wide range of scenarios including:

- **Happy Path:** Standard, valid inputs.

- **Edge Cases:** Boundary conditions (e.g., empty strings, zero, negative numbers).

- **Error Handling:** Cases that should raise specific exceptions.

Use the following format for your output:
\begin{lstlisting}[style=mypython]
```python
# your generated test code here
```
\end{lstlisting}

User Request:

\inputvar{\{user\_request\}}

Requirements

\inputvar{\{requirements\}}

Assumptions

\inputvar{\{assumptions\}}

Python function to test:

\inputvar{\{function\_to\_test\}}

Test cases:

\end{tcolorbox}

\begin{tcolorbox}[
  title=Prompt: Policy Code Correction,
  colback=PromptBackground,
  colframe=PromptFrame,
  fonttitle=\bfseries\color{PromptTitle},
  breakable, 
  arc=2mm,   
]
\newcommand{\promptsection}[1]{\par\medskip\textcolor{PromptSection}{\bfseries \#\# #1}\par\medskip}
\newcommand{\promptsubsection}[1]{\par\medskip\textcolor{PromptSection!80!black}{\bfseries \#\#\# #1}\par\medskip}
You are an expert Python developer and debugger. Your task is to analyze a Python function and its corresponding pytest error message, identify the bug, and provide the corrected code.

Python Function to Correct

\inputvar{\{function\_to\_test\}}

Pytest Error Message

\inputvar{\{error\_message\}}

\promptsection{Your Task}

Analyze the function and the error message to find the source of the error.

Explain the bug clearly and concisely.

Provide the complete, corrected Python function.

**Response Format**

Bug Explanation

(Describe the bug and the reason for the error here.)

Corrected Function

\begin{lstlisting}[style=mypython]
```python
# Your corrected Python code here.
```
\end{lstlisting}

\end{tcolorbox}

\subsection{Prompts Used for Verification}\label{ap:prompt_verification}
\begin{tcolorbox}[
  title=Prompt: Code Generation for Verification,
  colback=PromptBackground,
  colframe=PromptFrame,
  fonttitle=\bfseries\color{PromptTitle},
  breakable, 
  arc=2mm,   
]

\newcommand{\promptsection}[1]{\par\medskip\textcolor{PromptSection}{\bfseries \#\# #1}\par\medskip}

You are an expert in **formal methods** and **software verification**, specializing in Python. Your primary skill is translating requirements into precise **Nagini pre- and post-condition contracts**.

\promptsection{Objective:}
Your task is to augment a given Python function with Nagini contracts (`Requires` and `Ensures`) based on a set of logical statements. You must ensure the generated code is syntactically correct and accurately reflects the logic of the provided statements.

-----

\promptsection{Instructions}

1.  **Analyze the Inputs:** Carefully review the provided Python function and the list of requirements given as logical statements.

2.  **Translate Policies to Nagini:** For each logical statement, formulate the equivalent Nagini `Requires` (pre-conditions) or `Ensures` (post-conditions).

3.  **Adhere to Grammar:** Strictly follow the provided Nagini grammar and refer to the examples for correct syntax and structure.

4.  **Integrate and Output:** Embed the generated Nagini contracts directly into the Python function. 

-----

\promptsection{Inputs}
Python Function:
    
\inputvar{\{python\_function\_code\}}

Requirements:

\inputvar{\{list\_of\_logical\_statements\}}

Nagini Grammar Reference:

\inputvar{\{grammar\}}

Nagini Examples:

\inputvar{\{examples\}}

-----

\promptsection{Output Format}

Provide the complete Python code for the function, including the newly added Nagini decorators, inside a single Python code block.

\end{tcolorbox}

\end{document}